\documentclass[preprint,review,3p,times,twocolumn]{elsarticle}

\usepackage{amsmath}
\usepackage{amssymb}
\usepackage{amsthm}
\usepackage{hyperref}

\journal{Ledger Journal}

\begin{document}

\begin{frontmatter}



\title{Interactive Proof-of-stake}


\author{Alexander Chepurnoy}
\ead{kushti@protonmail.ch}
\address

\begin{abstract}
The paper examines decentralized cryptocurrency protocols that are based on the use of internal tokens as identity tools. An analysis of security problems with popular Proof-of-stake consensus protocols is provided. A new protocol, Interactive Proof-of-stake, is proposed. The main ideas of the protocol are to reduce a number of variables a miner can iterate over to a minimum and also to bring a communication into block generation. The protocol is checked against known attacks. It is shown that Interactive Proof-of-stake is more secure than current pure Proof-of-stake protocols. 
 \end{abstract}
 
\begin{keyword}
Cryptocurrency \sep Blockchain consensus protocols\sep Proof-of-stake \sep Peer-to-peer networks 
\end{keyword}

\end{frontmatter}


\section{Introduction}

The core feature of Bitcoin is is the ability of every network participant who contributes computational resources to participate in a shared ledger creation process. A few years after the Nakamoto's paper\cite{Nakamoto2008} described Proof-of-work and  private fork attack, some formal models with comprehensive analysis were published\cite{Garay2015,Miller2014a}.	

In a Proof-of-work system participants solve moderately hard computational puzzles\cite{Miller2014a} to generate valid blocks. The probability of generation is proportional to their computational power (with some exceptions\cite{Eyal2014,Garay2015}). The puzzles could be viewed as identity tools that prevent Sybil attacks in an anonymous environment. An alternative idea behind Proof-of-stake protocols uses internal tokens of a cryptocurrency as identity tools. Proof-of-stake was first implemented in PeerCoin\cite{King2012}. Then the concept was developed into several variations having similar problems with their security models.

NeuCoin\cite{Davarpanah2015} and Nxt are the Proof-of-stake protocol examples considered below.

\subsection{NeuCoin}

Neucoin uses the same transactional model as Bitcoin, with a transaction having multiple inputs and outputs. An unspent output holder has a right to generate a block if \(hash(k) \leqslant g*v_o*\delta\) where 
\(g\) is the \textit{target} value readjusted on each block and stored in a block header, \(v_o\) is the unspent output value, and \(\delta\) is the amount of time since the coins were touched. The \textit{kernel} \(k\) is:

\begin{equation}\label{eq:1} k = t \cup t_{utxo} \cup n_{utxo} \cup i_o \cup t_{block} \cup s\end{equation}

In the formula for \(k\), \(t\) is the current timestamp, \(t_{utxo}\) is a timestamp of the output, \(n_{utxo}\) is an index of the output in its transaction, \(i_o\) is an index of the output in its block, 
\(t_{block}\) is a timestamp of a block containing an output, and \(s\) is the \textit{stake modifier}, a 64-bit string seeded from the blockchain. 

\subsection{Nxt}

Nxt has a different transactional model than Bitcoin, with dedicated accounts holding balances. Accounts are allowed to generate a block if the \(hit\), an unsigned number constructed from 256-bit \textit{generation signature}, is less than the \textit{target}:

\begin{equation}f_{8bu}(g) < t*b*\delta \end{equation}

where \(f_{8bu}\) return the first 8 bytes of a byte array as an unsigned number, \(t\) is the \textit{base target} of a previous block, \(b\) is the account balance \(N\) blocks ago, and \(\delta\) is the delay in since the last block.

The \textit{generation signature} in the formula above is a deterministic value that depends on the value used to generate the previous block \(g_{prev}\) and the account's public key \(k_{pub}\):

\(g = hash(g_{prev} \cup k_{pub})\)

The \textit{base target} also changes between blocks. As the desirable delay between blocks is T seconds, the new base target can be calculated from the base target of a previous block \(t_{prev}\):

\(t = max(min(\frac{t_{prev}*\delta}{T}, t_{prev}*2), \frac{t_{prev}}{2})\)	

Every block header contains a generation signature \(g\) and a base target \(t\). \textit{Cumulative difficulty} \(d_c\) is used to determine the best blockchain, with larger values corresponding to better chains:

\begin{equation} d_c = \sum \frac {1}{t} \end{equation}

If every block in a chain is generated within \((\frac{T}{2},2*T)\) seconds of the previous block, \(d_c\) is proportional to the time elapsed since the genesis block timestamp.

\subsection{Why Proof-of-stake Matters}

There are several reasons to search for protocols beyond the well-known Bitcoin consensus protocols. First of all, Proof-of-stake provides an incentive to run a full-node\cite{Bentov2013}. Second, a new cryptocurrency using a Proof-of-work consensus protocol could be destroyed by Goldfinger attack\cite{Bonneau2015}. Proof-of-work could also be infeasible for private blockchains. While traditional Byzantine Agreements\cite{Cachin2005} could work well for a networks consisting of a few banks, Proof-of-stake seems to be more suitable for large private blockchains with unequal generation rights. 

\subsection{Structure of the Paper}

Section 2 describes some problems with existing Proof-of-stake consensus protocols. An alternative stake-based protocol, the Interactive Proof-of-stake, is proposed in Section 3. The novel approach is discussed in Section 4.

\section{Open Problems With Existing Proof-of-stake Implementations}
\label{s_problems}

Many attacks against Proof-of-stake consensus protocols have been proposed. Following list has been compiled based on papers\cite{Bentov2014,Bentov2013}, internet resources\cite{Buterin,Buterina,Poelstra} and private conversations.

\begin{enumerate}[1.]
\item Grinding Attacks

Iterations over some of the parameters in the formulas given in Sections 1.1 and 1.2 create the possibility for several attacks. In particular, as there are no precise global clocks in a distributed network an attacker can iterate over current time values to find a better option than a real timestamp of the local system. In addition, attackers can iterate over the public keys (for Nxt) or unspent output parameters (Neucoin). 

For example, with Neucoin's dynamic kernels it is possible to find a better kernel value by iterating over the allowed range of current timestamps. As NeuCoin allows block time to be set up to 2 hours later than a previous time, a selfish miner could calculate 7200 kernel values and immediately publish block with a timestamp from the future. Assuming that each network participant is selfish, the network could publish all of the blocks extending current blockchain within the next 2 hours, causing trivial system instability. 

Nxt is free from time drifting as hit value is static. However, it is possible to iterate over \(\delta\). To prevent this, Nxt has a 15 second propagation rule: incoming blocks are propagated only if their timestamps are within 15 seconds of that of a last local block. This limits the iteration range to at most 15 values.

\item Private-fork and Nothing-at-Stake attacks

Compared to Proof-of-work protocols, it is extremely cheap to start a fork with Proof-of-stake. Simulations show that short private forks are profitable for big stakeholders of Nxt\cite{Andruiman2014}. Therefore, selfish miners have an incentive to build on top of every fork they encounter. However, as the number of forks increases exponentially with time, it is not viable to contribute to all of them, so a node retains best forks. 

With the network contributing to a tree that contains \(F\) widely recognized forks Nothing-at-Stake attacks are emerging. This new kind of attack is made by a peer who is willing to double-spend by voting for an only chain containing a spending transaction (whereas the rest of the network works on all \(F\) forks). After some number of confirmations, a transaction recipient performs corresponding actions(for example, sending out goods). The attacker therefore contributes to \(F-1\) forks that do not contain a transaction. Simulations of Nxt-like consensus protocols with explicit tree support show the attack could be eliminated by an increase in the number of confirmations required before a transaction is considered as non-reversible\cite{Andruiman2015}.

\item History Attacks

Attackers can buy unused private keys held a majority of stake being online in the past, then a better chain could be generated \cite{Buterina}. A checkpointing mechanism is needed to prevent this kind of attack.

\item Bribe Attacks

After an attacker sends a transaction to the network, and some number \(k\) of confirmations a transaction recipient performs a corresponding action, such as sending goods out. Then a sender can publicly announce a reward for a better fork, reversing the last \(k+1\) blocks to remove the transaction from the blockchain. Such an attack is also possible with Proof-of-work protocol, but unsuccessful attack miners would lost a lot of resources in the attack. In contrast, bribed miners in a Proof-of-stake protocol will never lose anything because they can contribute to both forks for very little\cite{Bentov2014}.

\end{enumerate}

\subsection{Why Proof-of-stake Cryptocurrencies Work}

No practical implementations of the aforementioned attacks are currently known. There are two reasons for this:

\begin{enumerate}

\item Security Through Default Implementation

Nxt and private forks are discussed here. The single existing Nxt protocol implementation, Nxt Reference Software(NRS), stores only a single blockchain. A big stakeholder can earn more\cite{Andruiman2014} by adding an explicit local blocktree storage to the software. Then she can distribute the modified software. Nothing-at-Stake attacks are only a threat when a majority of online are using the modified software. There is a little incentive to modify a single copy even because current mining rewards are small. 
		
\item Security Through Checkpoints

With NRS, Nxt has a few hard-coded checkpoints and a 720-block reversal limit so only new nodes that are downloading the best chain from the genesis block could suffer a successful History attack resulting in a fork not deeper than a hardcoded checkpoint.

\end{enumerate}

Both practical security measures are inappropriate for long-term use with decentralized cryptocurrency that is intended for global adoption. In contrast, private blockchains can be assumed to be a running specific protocol implementation on most of their nodes. 

\section{Interactive Proof-of-stake}
\label{s_ipos}

There are two main idea behind this new protocol. First of all, a number of variables a miner can iterate over is needed to be reduced to a minimum to increase resistance to grinding attacks. Second, we would like to bring the communication process into block generation. It requires multiple parties for block creation without broadcasting undersigned blocks within the network. The proposal for this protocol is as follows:

\begin{enumerate}

\item The presence of accounts with non-zero is assumed. Accounts can be viewed as \textit{(public key, balance)} pair. For monetary blockchains we can consider monetary tokens to be balance units. For private non-monetary blockchains we can set correspondences in a genesis block so that network participants have transferable or non-transferable \textit{generation rights}.

\item Block headers contain some unique \(seed\) value known to all participants to determine the generators of the next blocks.

\item \(T\) accounts are needed to generate a single block. We consider \(T = 3\). To avoid network propagation of block candidates do not have enough signatures, only one account is allowed to sign and broadcast a block, but each block must contain \(T\) \textit{tickets}. For \(T = 3\) we denote ticket types as \(Ticket1\), \(Ticket2\), \(Ticket3\). To generate a block one instance of each type is necessary.  

\item Each ticket is made of a corresponding block \(seed\), a public key \(pk\) and the balance \(b\) of an account. Using this data we can calculate a \textit{score} \(s_t\) of a ticket. First, we calculate \(hash(seed \cup pk)\). Then the first byte of the digest is used to generate \(Ticket1\), the second, to generate \(Ticket2\), and the third to generate \(Ticket3\). The byte value is an unsigned number \(m\). If  \(0 < m \leqslant R\), where \(R\) is some constant, then \(s_t=m * \log_2 b\); otherwise \(s_t=0\).

\item For a block of height \(h\) an account can generate up to three tickets with positive score: an instance of \(Ticket1\) by using the \(seed\) of a block at height \(h-2\), an instance of \(Ticket2\), using the \(seed\) of a block at height \(h-1\), and an instance of \(Ticket3\), using the \(seed\) of a block at height \(h\). If a ticket's score is positive, then the account signs it and broadcasts to the network. Ticket size (including a signature) is about 100 bytes, so tickets can be propagated around the network quickly and effectively. 

\item Only one ticket per account per last \(l\) blocks is allowed.

\item Only the \(Ticket1\) instance generator forms a block containing transactions and the three tickets.

\item A blockchain begins with three genesis blocks instead of one to avoid breaking ticket generation rules for the first block after genesis. 

\item If the three tickets of a block are generated by accounts with public keys \(pk_{1}\), \(pk_{2}\), and \(pk_{3}\) then the \(seed\) of that block is calculated as \(seed = hash(seed_{prev} \cup pk_{1} \cup pk_{2} \cup pk_{3})\), where \(seed_{prev}\) is the seed of a previous block.	

\item A block's score \(s_b\) is the sum of its ticket scores: \(s_b = s_{t_1} + s_{t_2} + s_{t_3}\). A blockchain's score is the sum of its block scores: \(s_{bc}=\sum{s_b}\). The chain with the greatest score wins. Block generators therefore have an incentive to build blocks using the best tickets available.

\item If \(R < 255 \), then it is possible for all miners to generate \(m\) values outside the range, meaning no tickets will be generated. Some protection against that is required for pure Proof-of-stake systems, for example, if the delay since the last block is more than \(D\) seconds then for a new block \(R = 255, s_{t_i} = m\). If \(D\) is large, a chain without hanging problems will get a block with a better \(s_{t_i}\) values, and additional blocks will be generated during the delay.

\item Block reward is to be split equally amongst ticket generators.

\item It is necessary to limit block generation frequency. For example, a minimum delay after the last block and a propagation rule like the one used for Nxt could be specified. 

\end{enumerate}

Consider a scenario in which \(N_A\) accounts are online. The ticket generation could be seen as a two-step weighted lottery. First, a random subset of average size \(\frac{(R-1)*N_A}{256}\) will be chosen regardless on the account stake. Then each account in the set will obtain some random number \(m\) (\(0 < m \leqslant R\)) to generate a ticket with a score \(m * \log _2 b\). 

Considering the moment at which block \(b_h\) (of height \(h\)) arrives at a node that has an account \(A\) onboard. \(A\) can generate up to three tickets with positive scores: an instance of \(Ticket3\) for a block of height \(h+1\), an instance of \(Ticket2\) for a block of height \(h+2\), and an instance of \(Ticket1\) for a block of height \(h+3\). Tickets with positive scores will be broadcasted immediately. At the moment of arrival, \(A\) supposedly also has tickets based on blocks \(b_{h-3}\), \(b_{h-2}\), and \(b_{h-1}\). In particular, \(A\) could have a better ticket than any of \(b_h\)'s. In that case it is reasonable to postpone \(b_h\) processing and wait for a better block for some time. If \(A\) can generate a better block than \(b_h\), \(A\) will do so and broadcast that block instead of \(b_h\) processing.

\section{Discussion of the Protocol}
\label{s_discuss}

\subsection{Time and Balance}

The protocol does not include timestamps or time delays and is therefore immune to time drifting attacks. The protocol's balance is static. To implement the protocol, we could use an account balance from \(N\) blocks ago, as Nxt does.

\subsection{The Protocol Simulation}

An executable simulation of the protocol has been published online under a public domain license \cite{Chepurnoy}.

Wealth distribution in Bitcoin is described by a stretched exponential function\cite{Kondor2014}. For simplicity, a negative exponential distribution with \(\lambda=50\) was used to simulate block generation. The simulation consisted of 800 accounts sharing approximately 1.6 billion coins and generating 30,000 empty blocks (\(R=16\), \(l=10\)). The results were as follows: 

\begin{enumerate}
\item \(\approx 89\% \) of accounts generated at least one ticket

\item Poor(\(< 0.1\%\) of stake) and rich(\(>0.7\%\)) accounts held disproportionally low shares of tickets
\end{enumerate}

\scalebox{0.45}{\includegraphics{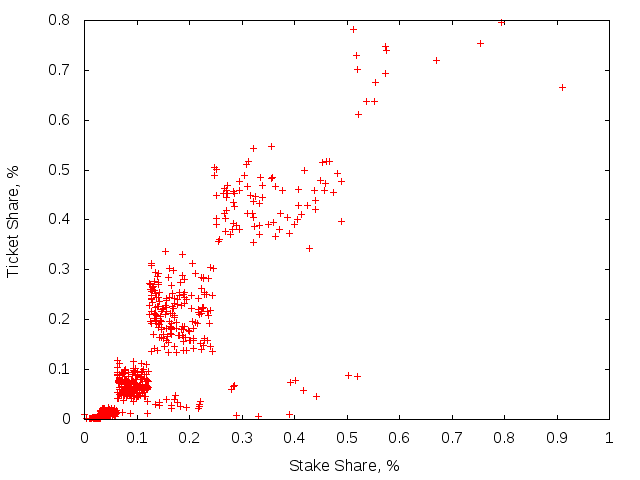}}

\subsection{Resistance to Attacks}

\begin{enumerate}

\item Private forks and History attacks

Even very rich single account can generate no more than one ticket per \(l\) blocks. To maximize her chance of lottery participation, a large stakeholder needs to split her funds into \(\gg \frac{256}{R+1}\) accounts. An attacker's accounts have a better chance to generate a better fork if for a majority of possible clusters of size \(\frac{R}{256+1}\) an account controlled by her has a greater balance than others. Therefore, to maximize a chance of an attack success on a cryptocurrency with exponential distribution of wealth the best strategy is to split stake equally into as small as possible number of parts while enough to generate all the tickets with no hanging. The tool for simulating this kind of attack is published online\cite{Chepurnoy}. We simulated a competition between a network of 800 accounts and an attacker. The network had an exponential wealth distribution (\(\lambda = 50\)) and approximately 1.6 billion coins.  The attacker had \(x\) percent of a network stake, equally divided into \(P\) parts. For \(R=16, l=10, P \approx 180\) is the best choice; an attacker with \(x=50\) has \(\approx 8.3\%\) chance of generating a better chain of length 10 and no chance of generating a better chain of length 50. This means that a network with the given parameters is secure against private fork attack by richest holders of \(\frac{1}{3}\) of online stake, for 50 confirmations. The same adversarial power against Nxt consensus protocol has \(\approx 14\%\) chance of generating a better chain of length 10 and \(\approx 1.7\%\) chance of generating a better chain of length 50. 

Under conditions of optimal stake distribution, History attacks have the same chances of success as the aforementioned private fork attacks if none of the purchased accounts participate in block generation after the attack has begun. In practice, both requirements are unlikely to be met, so the chance of success will be lower than of a private fork attack.

\item Bribe attacks

We consider a set of online accounts to be static during a bribe attack. At the moment of bribe proposal the best possible chain suffix of \(k\) blocks has already been generated and therefore cannot be substituted. An attacker can offer a bribe in advance of an attack to postpone best ticket publishing, but the ticket reward will be lost if the attack fails. Also collusion between more parties is required than in existing Proof-of-stake protocols.

\item Grinding and Iteration attacks

Forcing block seeds to choose more attacker accounts in the future is the only method of attack. As block seeds depend on the public keys of all three ticket generators, an attacker's accounts need to generate more than three best possible tickets for a block to iterate over them. Thus grinding attacks are harder to perform in comparison with Nxt and Neucoin. 

\end{enumerate}

\subsection{Block Trees}

We argue there is no incentive even for a large stakeholder to modify the node software aiming to contribute to multiple chains. Even if she has enough stake to compete with the rest of the network, she is whether working on her own chain or network's. However, if the majority of miners are already working on multiple chains, the best strategy for a single miner is to contribute to all of them. 

\section{Related Work}
\label{s_related}

The prospect of having multiple block generators was raised in the Proof-of-activity proposal\cite{Bentov2013}, which considered extending Proof-of-work via additional stakeholder signatures. The Chains-of-activity proposal\cite{Bentov2014} extends the Proof-of-activity lottery to a pure Proof-of-stake protocol, in which each block provides a single random bit and a sequence of random bits determines a future block generator. The Tendermint whitepaper\cite{Kwon} offers Byzantine Agreements to participants who have made security deposits.

\section{Further Work}
\label{s_further}

\subsection{Dynamic \(R\) Calculation}

It can be difficult to set the best single static \(R\) value for the entire lifespan of a blockchain system. The more sophisticated option is to dynamically change \(R\) based on generation process characteristics. This could make private fork attacks harder to perform.

\subsection{Proof-of-w ork Tickets}

Tickets could be used in Proof-of-Work protocols aiming to spread rewards more evenly and reduce selfish mining attack\cite{Eyal2014} probability.

\subsection{A Hybrid Protocol}

The Interactive Proof-of-stake protocol can be used in combination with Proof-of-work. Empty Proof-of-work blocks could act as random beacons\cite{Bonneau2014} to generate the tickets for the next \(S\) blocks. They could also act as decentralized voting tools to enforce chain selection in a Proof-of-stake blocktree. External random beacons could eliminate grinding atacks. Chain selection enforcement for a blocktree via decentralized voting could eliminate forks deeper than \(S\) blocks. A hybrid protocol would remind Bitcoin-NG\cite{Renesse} with many stakeholders (rather than a single miner) working on microblocks. This way the network participants have Proof-of-work security guarantees, fast blocks and an incentive to run a full-node.

\section{Conclusion}
\label{s_conclusion}

We have analyzed possible security problems of current Proof-of-stake blockchain consensus protocols compiling a comprehensive list of attacks known at the moment of writing the paper. A new pure Proof-of-stake protocol with multiple block generators, Interactive Proof-os-stake, is proposed. It has a minimum number of variables a miner can iterate over. It has no timestamps or time delays explicitly stated. Instead, the new protocol operates as a weighted lottery where multiple winners create a block. We have checked the protocol against known attacks list. Results show that a global cryptocurrency running the protocol is immune to a private fork attack made with up to \(\frac{1}{3}\) of network's online stake(for 50 confirmations); a bribe attack becomes more tricky and requires more parties to collude with a participant losing a ticket reward in case of failed attack; grinding attacks are possible over public keys only and harder to perform than in Nxt and NeuCoin. However, we do not claim the security of our protocol is better or the same than that of Proof-of-work protocols. Nevertheless, the new protocol could be useful for private blockchains and hybrid (Proof-of-work mixed with Proof-of-stake) protocols.

\section*{Acknowledgement}

I would like to thank Benjamin Cordez, Charles Hoskinson, Serguei Popov(a.k.a. mthcl), Sergey K. and Dmitry M. for their feedback.

\section*{Author Contributions}

Alexander Chepurnoy proposed the protocol, did the paper and simulation tools. 

\section*{Conflict of Interest}

The author used to be an Nxt core developer since April, 2014 until May, 2015. That should not be considered as the author is biased positively or negatively towards Nxt or any competitor of it at the moment of writing the paper(December, 2015).

\section*{References}

\bibliographystyle{elsarticle-num}
\bibliography{sources.bib}

\end{document}